\documentclass[12pt]{article}
\usepackage{graphicx} 
\newcommand\slv{v\kern-5pt\raise1pt\hbox{$\scriptstyle/$}\kern1pt}

\begin{document}
\thispagestyle{empty}
\begin{flushright}
WUE-ITP-99-011\\
\end{flushright}
\vspace{0.5cm}
\begin{center}
{\Large \bf Perturbative Effects in the Form Factor 
$\gamma\gamma^*\to \pi^0$}\\[.3cm]
{\Large \bf  and Extraction of the Pion Distribution Amplitude}\\[.3cm]
{\Large \bf from CLEO Data}\\

\vspace{1.7cm}
{\sc A.~Schmedding$^{1,a}$ and  O.~Yakovlev$^{1,2,b}$\\[1cm]
$^1$ {\em Institut f\"ur Theoretische Physik, Universit\"at W\"urzburg,
D-97074 W\"urzburg, Germany} \\
\vspace{4mm}
$^2$  {\em Randall Laboratory of Physics, University of Michigan,\\ 
Ann Arbor, Michigan 48109-1120, USA} 
}
\end{center}
\vspace{2cm}
\begin{abstract}\noindent

We study the pion form factor $F^{\pi \gamma\gamma^*}(Q^2)$ 
in the light-cone sum rule approach, accounting for radiative 
corrections and higher twist effects. 
 Comparing the results to the CLEO experimental data on 
$F^{\pi \gamma\gamma^*}(Q^2)$, we extract the the pion 
distribution amplitude of twist-2. 
The deviation of the distribution amplitude from the asymptotic 
one is small and is estimated to be $a_2(\mu) = 0.12 \pm 0.03 $ 
at $\mu=2.4$ GeV, in the model with one non-asymptotic term.   

The ansatz with two non-asymptotic terms gives some region 
of $a_2$ and $a_4$, which is consistent with the asymptotic 
distribution amplitude, 
but does not agree with some old models. 

\end{abstract}

\centerline{\em PACS numbers: 11.55.Hx, 12.38.Bx, 13.40.Gp, 14.40.Aq}
\vspace{1cm}

\vspace*{\fill}

\noindent $^a${\small e-mail: schmedding@physik.uni-wuerzburg.de}\\
\noindent $^b${\small e-mail: yakovlev@umich.edu}

\newpage
\section {Introduction}
The production process of one neutral pion by  
two virtual photons, $\gamma^*\gamma^*\to\pi^0$, plays a crucial 
role in the studies of exclusive processes in
quantum chromodynamics. 
Being one of the simplest exclusive processes, it involves only 
one hadron and relates directly to the pion distribution amplitude \cite{BL}.
At large photon virtualities, we can calculate the form factor 
using perturbative QCD and obtain important 
information on the shape of the pion distribution amplitude
from the experimental data.

In general, the pion distribution amplitudes serve as input in the QCD 
sum rule method and  allow the calculation of many form factors 
(for example heavy-to-light form  factors $B\to\pi$ and  $D \to \pi $) 
and  hadronic coupling constants (for example $g_{B^*B\pi}$ and  
$g_{D^*D\pi}$).  We refer the reader to reviews 
\cite{Braun, RKrev} and recent studies \cite{B2piQCD}. 
   
Recently, the CLEO collaboration has measured the $\gamma\gamma^*\to \pi^0$ 
form factor. In this experiment, one of the photons 
is nearly on-shell and the other one is highly 
off-shell, with a virtuality in the range $1.5$ GeV$^2$ - $9.2$ 
GeV$^2$ \cite{CLEO} \footnote{ There exist also older results by 
the CELLO collaboration \cite{CELLO}.}.  
We study the possibility of extracting 
the twist-2 pion distribution amplitude from the CLEO data. 

The pion-photon transition has been the subject 
of many studies in framework perturbative QCD \cite{RadMus}, 
lattice calculations \cite{Latt} and 
QCD sum rule method \cite{SR,RadRus,RadMus,AK}. 

In the present paper, we use the light-cone sum rule (LCSR) method
for calculating the form factor of the process 
$\gamma\gamma^*\to \pi^0$ .
The method of LCSR  has been suggested in \cite{BraunLC,CZhLC} and 
consists of the operator product expansion (OPE) on the 
light-cone \cite{CZ,ER,BL} combined with the QCD 
sum rule technique \cite{SR}. 
 
The first attempt to calculate this form factor by using the LCSR
method has been reported in \cite{AK}.   
The twist-2 and twist-4 contributions were calculated 
to leading order without accounting for perturbative QCD effects.
The radiative QCD effects are usually large ($20\%$) and 
 allow to fix the normalization scale dependence of involved parameters.

In this  paper we analyze the LCSR for the form factor  
$\gamma\gamma^*\to \pi^0$ at next-to-leading order in $\alpha_S$ (NLO). 
We derive ${\mathcal O}(\alpha_S)$ radiative corrections to the 
spectral density of the twist-2 operator. 
We combine the twist-2 contribution at NLO 
with higher twist contributions
(twist-4) in order to analyze the LCSR for 
the form factor of the process $\gamma\gamma^*\to \pi^0$  numerically. 

 Using CLEO experimental data, we extract 
the parameters of the twist-2 distribution amplitude.
The distribution amplitude appears to be very close to 
the asymptotic distribution amplitude.     
 The deviation of the distribution amplitude from the asymptotic 
one is small and is estimated to be 
$$a_2(\mu) = 0.12 \pm 0.03\quad\mbox{at}\quad \mu=2.4\quad\mbox{GeV},$$ 
using pion distribution amplitude with one non-asymptotic term. 
This result on $a_2$ agrees well with recent 
analysis on electromagnetic pion form factor \cite{BKM}.
 
The pion distribution amplitude with two non-asymptotic terms 
is also extracted, giving some region in the $a_2,a_4$-plane.
 Theoretical systematical uncertainties are dominant over 
experimental statistical ones and systematical uncertainties 
are strongly correlated, defining the allowed parameter space of $a_2,
a_4$. The extracted region for the distribution amplitude
is in qualitative agreement with results derived in  
\cite{MikhRad,BakMikh,Belyaev,BKM}, where it has been claimed
that the pion distribution amplitude is very close to the asymptotic
form.  However, our region does not overlap 
with the pion distribution amplitudes suggested in \cite{BF} and 
\cite{CZ}. 

The paper is organized as follows. 
In the next section we discuss the general framework  of  calculating  
the $\pi\gamma\gamma^*$ form factor with LCSR. 
In section 3 we calculate a spectral density at LO and 
NLO and present the final sum rule at NLO. 
In  section 4 we perform a numerical analysis and discuss 
a procedure to extract the parameters 
$ a_2 $ and $ a_4$ including the estimation of the systematic 
and statistical uncertainties. 

\section{The method of calculation}

 We start with the correlator of two vector currents 
$j_{\mu/\nu}=(\frac{2}{3}\bar u \gamma_{\mu/\nu} u 
-\frac{1}{3}\bar d \gamma_{\mu/\nu} d )$:
\begin{eqnarray}\label{Correlator}
\int d^4 x e^{-iq_1x}<\pi^0(0)|T\{j_\mu(x) j_\nu(0) \}|0>=
i\epsilon_{\mu\nu\alpha\beta}q_1^\alpha q_2^\beta 
F^{\pi\gamma^*\gamma^*}(q_1,q_2),
\end{eqnarray}
where $q_1, q_2$ are the momenta of the photons. 
If both virtualities $s_1=q_1^2$ and $s_2=q_2^2$ 
are large and Euclidean, $-s_1,-s_2 \gg \Lambda_{QCD}$, 
the correlator can be expanded near the light-cone ($x^2\to 0$). 
The leading  twist-2 contribution to the correlator 
(\ref{Correlator}) is \cite{BL}
\begin{eqnarray}\label{TW2as}
F^{\pi\gamma^*\gamma^*}(s_1,s_2)=\frac{\sqrt{2}f_\pi}{3}\int\limits_{0}^{1}
\frac{du\varphi_\pi (u)}{s_2(u-1)-s_1u},
\end{eqnarray}
where $\varphi_\pi (u)$ is the pion distribution amplitude of twist-2 defined through
\begin{eqnarray}\label{PionME}
\langle \pi(q)|\bar q(x)\, \gamma_\mu\, \gamma_5\, q(0)|0\rangle = 
-i\, q_\mu\, f_\pi\,\int\limits_0^1\!\! du\,\varphi_\pi(u)\,e^{i\,uqx}.
\end{eqnarray}
In principle, the higher twist contributions can be 
calculated using the OPE on the light cone.

However, in the CLEO experimental data, 
one of the virtualities is small, i.e. $s_2\to 0$.
A straightforward  OPE  calculation is not possible
and we have to use analyticity and duality arguments.

Since the form factor $F^{\pi\gamma^*\gamma^*}(s_1,s_2)$ is an analytical 
function in both variables, we can write the form factor 
as a dispersion relation in $s_2$:
\begin{equation}\label{E:hr}
F^{\pi\gamma^*\gamma^*}(s_1,s_2) = \frac{\sqrt{2}\,f_\rho\,F^{\rho\pi}(s_1)}
 {m_\rho^2-s_2} + \int\limits_{s_0}^\infty\!\!ds\,\frac{\rho^h(s_1,s)}
 {s-s_2}.
\end{equation}
The physical ground states are  $\rho$ and $\omega$ vector mesons.
We use the zero-width approximation and define the matrix elements of
electromagnetic currents, assuming isospin symmetry:
$m_\rho\simeq m_\omega$;   
$\frac{1}{3}\langle \pi^0(p)|j_\mu|\omega(q_2)\rangle \simeq
 \langle \pi^0(p)|j_\mu|\rho^0(q_2)\rangle = 
  \frac{1}{m_\rho}\,\epsilon_{\mu\nu\alpha\beta}\,e^\nu\,q_1^\alpha
  \,q_2^\beta\,F^{\rho\pi}(s_1)$;  
$3\,\langle\omega|j_\nu|0\rangle \simeq \langle\rho^0|j_\nu|0\rangle  =
  \frac{f_\rho}{\sqrt{2}}\,m_\rho\,e^*_\nu$. 
Here $e_{\nu}$ is the polarization vector of the $\rho$ meson; 
$f_\rho$ is its decay constant.

The spectral density  of higher energy states $\rho^h(s_1,s)$ is derived 
from the QCD-calculated expression for  $F_{QCD}^{\pi\gamma^*\gamma^*}(s_1,s)$
by using usual semi-local quark-hadron duality for  $s>s_0$ 
\begin{equation}\label{E:duality}
\int ds \mbox{(any function)} \rho^h(s_1,s) = 
\int ds \mbox{(any function)} 
\frac{1}{\pi}\,\mbox{Im}\,F_{QCD}^{\pi\gamma^*\gamma^*}(s_1,s).
\end{equation}

We equate the dispersion relation (\ref{E:hr}) 
with the QCD expression at large  $s_2$. 
Using the dispersion relation for the QCD function 
$F_{QCD}^{\pi\gamma^*\gamma^*}(s_1,s_2)$, we obtain
\begin{equation}
\frac{\sqrt{2}\,f_\rho\,F^{\rho\pi}(s_1)}
 {m_\rho^2-s_2} 
= \frac{1}{\pi}\int\limits_0^{s_0}ds\,
  \frac{\mbox{Im}\,F_{QCD}^{\pi\gamma^*\gamma^*}(s_1,s)}{s-s_2}.
\end{equation}
The next step is to perform a Borel transformation
in $s_2$. We finally  get the LCSR for the form factor 
$F^{\rho\pi}(s_1)$:
\begin{equation}\label{E:rho}
\sqrt{2}\,f_\rho\,F^{\rho\pi}(s_1) = \frac{1}{\pi}\,\int\limits_0^{s_0}\!\! ds\,
 \mbox{Im}\,F_{QCD}^{\pi\gamma^*\gamma^*}(s_1,s)\, 
  \mbox{e}^{\frac{m_\rho^2-s}{M^2}},
\end{equation}
where $M$ is a Borel parameter.

Substituting  (\ref{E:rho}) and the duality approximation (\ref{E:duality}) 
into (\ref{E:hr}) and taking the $s_2\to 0$ limit we obtain the sum rule for
the form factor  $F^{\gamma\gamma^*\pi}(s_1)$ \cite{AK}:
\begin{equation}\label{E:srggpi}
F^{\pi\gamma\gamma^*}(s_1) = \frac{1}{\pi\,m_\rho^2}\,\int\limits_0^{s_0}\!\!
ds\,\mbox{Im}\,F_{QCD}^{\pi\gamma^*\gamma^*}(s_1,s)\, 
  \mbox{e}^{\frac{m_\rho^2-s}{M^2}} + \frac{1}{\pi}\,\int\limits_{s_0}^\infty
  \!\!\frac{ds}{s}\,\mbox{Im}\,F_{QCD}^{\pi\gamma^*\gamma^*}(s_1,s).
\end{equation}
We will use this expression as our basic sum rule for the 
numerical analysis. 

\section{Born term and QCD radiative correction} 

The next step is to calculate the spectral density at LO and NLO.

Calculating  twist-2 contributions to 
$F^{\pi\gamma^*\gamma^*}(s_1,s_2)$, 
only one distribution amplitude enters, the pion distribution amplitude, 
$\varphi_\pi(u)$, defined by (\ref{PionME}).
As a result, the 
$F^{\pi\gamma^*\gamma^*}(s_1,s_2)$ 
can be written as the convolution of the hard amplitude $T(s_1,s_2,u)$
and the distribution amplitude $\varphi_\pi(u)$:
\begin{eqnarray}
F^{\pi\gamma^*\gamma^*}(s_1,s_2)=f_\pi\,\int\limits_0^1\!\! 
du\,\varphi_\pi (u)\, T(s_1,s_2,u).
\end{eqnarray}
The hard amplitude -- it plays the role of the 
Wilson coefficient in OPE -- 
is calculable within perturbative theory, the pion distribution amplitude
 $\varphi_\pi(u)$ contains the long-distance effects.

The theoretical spectral density at $s_2 >0$ and $s_1<0$ 
can be calculated from
\begin{eqnarray}
\rho_{QCD}(s_1,s_2)=\frac{1}{\pi}\mbox{Im}_{s_2}
F^{\pi\gamma^*\gamma^*}(s_1,s_2)=f_\pi\,\int\limits_0^1\!\! 
du\,\varphi_\pi (u)\, \frac{1}{\pi} \mbox{Im}_{s_2} T(s_1,s_2,u).
\end{eqnarray}

The contribution of twist-2 at LO approximation (\ref{TW2as}) \cite{BL} is
\begin{eqnarray}
T_{QCD}^{LO}(s_1,s_2,u)=\frac{\sqrt{2}}{3}\frac{1}{s_2(u-1)-s_1u}\quad 
\mbox{and} \quad \rho^{LO}_{QCD}=\frac{\sqrt{2}f_{\pi}}{3(s_2-s_1)} 
\varphi_\pi (u_0).
\end{eqnarray}
where $u_0=\frac{s_2}{s_2-s_1}.$
 
The ${\mathcal O}(\alpha_s)$ correlation function for 
$\gamma^*\gamma^* \to \pi^0$ was considered
in \cite{Braaten} (see also \cite{Chase,KMR}). 
Here we use this result and present it 
in a form which is useful for our further calculations 
($c=\frac{\sqrt{2}f_{\pi}}{3(s_2-s_1)} \frac{\alpha_s (\mu) C_F}{2\pi}$):
\begin{eqnarray}
T_{QCD}=T_{QCD}^{LO}+T_{QCD}^{NLO};
\end{eqnarray}
\begin{eqnarray}\label{TT}
T^{NLO}_{QCD}(s_1,s_2,u) =c\left( a_0\,L_0 + a_1\,L_1 +
a_2\,L_2\right),
\;\;\mbox{with } L_n = \frac{\log^n(u-u_0)}{u-u_0}.
\end{eqnarray}
The coefficients $a_0$ are expanded in terms of $\log^n(-u_0)$:
\begin{equation}
a_0 = b_{0}\,l_0 + b_{1}\,l_1 + b_{2}\,l_2
\;\;\mbox{with } l_n = \log^n(-u_0).
\end{equation}
The coefficients are:
\begin{eqnarray}\label{EQ:coefficients}
b_{2} &=& \frac{-s_2^2\,u + s_2\,s_1(1+u)}{2u(s_2-s_1)^2},
\\\nonumber
b_{1} &=& -2\,b_{2}\,(L_\mu+1) - \frac{s_2}{2\,(s_2-s_1)\,u},
\\\nonumber
b_{0} &=& L_u^2\,b + L_u\left(\frac{s_1}{2\,(s_2-s_1)(u-1)} -
2\,(L_\mu+1)\,b\right)-\frac{3}{2}(3+L_\mu),
\\\nonumber
a_{2} &=& \frac{1}{2} + \frac{s_2\,s_1}{2(s_2-s_1)^2u(u-1)},
\\\nonumber
a_{1} &=&  -2\,a_{2}\,L_\mu + \frac{s_2^2\,(1-u)^2 + s_1^2u^2 +
s_2\,s_1\,(-3+2u-2u^2)}{2\,(s_2-s_1)^2(u-1)\,u},
\end{eqnarray}
with 
$$
L_\mu=\log(\frac{\mu^2}{s_2-s_1});\;\;\;\;
L_u = \log(1-u_0);\;\;\;\;
b = \frac{s_1^2\,(1-u) + s_2\,s_1\,(u-2)}{2\,(s_2-s_1)^2(u-1)}.
$$

Now we take the imaginary part of this expression in the energy region 
$s_1<0$ and $s_2>0$. The nontrivial imaginary part comes from 
the functions $L_n$ and $l_n$. We collect all useful formulae in Appendix A. 
The combined result is

\begin{eqnarray}\label{ImTT}
\frac{1}{\pi}\!\!\!\!\!\!\!\!&&\!\!\!\!\!\mbox{Im}\,
T^{NLO}_{QCD}(s_1,s_2,u)
=-c\Bigg( \mbox{P}\left(\frac{1}{u-u_0}\right)(b_{1}+2\,\log(u_0)\,b_{2})
\\\nonumber
&-&\delta(u-u_0)\left( 
b_{0}+ \log(u_0) (b_{1}+a_{1}) + \log^2(u_0)(b_{2}+a_{2}) 
-\pi^2\left(b_{2}+\frac{1}{3}a_{2}\right)\right)
\\\nonumber
&+&\Theta(u_0-u)\left(a_1\left(\frac{1}{u-u_0}\right)_+ +
a_2\left(\frac{2\log|u_0-u|}{u-u_0}\right)_+\right) \Bigg),
\end{eqnarray}
where we use  the $()_+$ operation, which is defined inside the integral as:
$$
\int du\,f(u)\Big( g(u,u_0)\Big) _+ = \int du\,\Big( f(u) - f(u_0)\Big)
g(u,u_0).
$$
We have checked that the dispersion integral with the 
imaginary part (\ref{ImTT}) reproduces the hard amplitude (\ref{TT}).

The expression (\ref{ImTT}) is  universal and could be used with any 
distribution amplitude. 
In the present paper we expand  the distribution amplitude 
in Gegenbauer polynomials, keeping the first three terms 
(see also next section for details):
\begin{equation}
\varphi_\pi^{tw2}(u,\mu) = 6\,u(1-u)\left(1+a_2(\mu)\,C_2^{3/2}(2\,u-1)+
  a_4(\mu)\,C_4^{3/2}(2\,u-1)+\ldots\right).
\end{equation}
The direct integration over variable $u$ gives the theoretical spectral 
density
\begin{eqnarray}\label{spectralNLO}
\rho^{NLO}_{QCD}(s_2,s_1,\mu)&=&
\frac{\sqrt{2}f_{\pi}}{3} \frac{\alpha_s (\mu) C_F}{2\pi}
  \,\Big( {A_0} (s_2,s_1) \\\nonumber
&+& a_2(\mu)\,{A_2}(s_2,s_1,\mu) +
    a_4(\mu)\,{A_4}(s_2,s_1,\mu)\Big),
\end{eqnarray}
with the coefficients
\begin{eqnarray}\label{coeff}
{A_0}&=&
{\frac{ s_2\,
       s_1\,}{{{\left( s_2 - s_1 \right) }^3}}
       \left( -15 + {{\pi }^2} - 
         3\,{{\log (-{\frac{s_2}{s_1}})}^2} \right) 
         },\\[3mm]\nonumber
{A_2}&=&-
\frac{\,s_2}{4\,
     {{\left( s_2 - s_1 \right) }^5}}\,
     \bigg( -25\,{{s_2}^3} - 
       8\,\left( 95 + 3\,{{\pi }^2} \right) \,{{s_2}^2}\,
        s_1 - 36\,\left( 25 + 2\,{{\pi }^2} \right) \,
        s_2\,{{s_1}^2} \\\nonumber&-& 
       12\,\left( 5 + 2\,{{\pi }^2} \right) \,{{s_1}^3} 
       + 
       12\,s_1\,\left( {{s_2}^2} + 
          3\,s_2\,s_1 + {{s_1}^2} \right) \,
        \left( -25\,\log ({\frac{{{\mu}^2}}{s_2}}) 
             + 6\,{\log (-{\frac{s_2}{s_1}})}^2
                \right)  \bigg) ,\\[3mm]\nonumber
{A_4}&=&-
\frac{s_2}{10\,
     {{\left( s_2 - s_1 \right) }^7}}\,
     \bigg( -91\,{{s_2}^5} - 
       2\,\left( 5413 + 75\,{{\pi }^2} \right) \,{{s_2}^4}\,
        s_1 - 125\,\left( 541 + 12\,{{\pi }^2} \right) \,
        {{s_2}^3}\,{{s_1}^2} \\\nonumber&-& 
       100\,\left( 901 + 30\,{{\pi }^2} \right) \,{{s_2}^2}\,
        {{s_1}^3} - 150\,\left( 193 + 10\,{{\pi }^2} \right) \,
        s_2\,{{s_1}^4} - 
       15\,\left( 109 + 10\,{{\pi }^2} \right) \,{{s_1}^5} \\\nonumber
       &+& 
       30\,s_1\,\left( {{s_2}^4} + 
          10\,{{s_2}^3}\,s_1 + 
          20\,{{s_2}^2}\,{{s_1}^2} + 
          10\,s_2\,{{s_1}^3} + {{s_1}^4}
           \right) \times\\\nonumber
           &&\left( -91\,\log ({\frac{{{\mu}^2}}{s_2}}) 
             + 15\,{\log (-{\frac{s_2}
                  {s_1}})}^2   \right)  \bigg). 
\end{eqnarray}
As a result the asymptotic contribution 
of twist-2 operators at NLO has a very simple form
\begin{eqnarray}
\rho_{QCD}=
\frac{2\sqrt{2}f_\pi s_2s_1}{\left( s_2 - s_1 \right)^3}
       \Big( 1+\frac{\alpha_s(\mu)C_F}{12\pi}\Big( -15 + \pi ^2 - 
         3\log (-\frac{s_2}{s_1})^2 \Big)\Big)
\end{eqnarray}
We note here that the spectral density contains double logarithms
$\alpha_s \log (-\frac{s_2}{s_1})^2$. Numerically, these are moderate 
in our LCSR with $s_2\approx M^2 \approx s_0 = {\mathcal O}(1\,\mbox{GeV}^2)$
 and $-s_1< 10$ GeV$^2$ 
(as in the case of CLEO data).
For higher virtualities, $-s_1\gg 10$ GeV$^2$, the resummation of 
the double logarithms will be necessary. 

 Finally, we obtain LCSR by combining the general expression (\ref{E:srggpi})
with the formulae for the twist-2 spectral density at NLO  derived 
in this paper (\ref{ImTT}, \ref{spectralNLO}, \ref{coeff})
with the twist-4 contribution taken from the literature \cite{AK}.

\section{Numerical results}

We use the following parameters from the Particle Data Group \cite{PDG}
 for the numerical analysis: 
$f_\pi=132\,\mbox{MeV}$, $f_\rho=216\,\mbox{MeV}$ and $m_\rho=770\,\mbox{MeV}$
\cite{PDG}. 
For the running of the QCD coupling constant $\alpha_S$, 
we use the two-loop expression 
with $N_f=3$ and  $\Lambda^{(4)}=380\, \mbox{MeV}$
which corresponds to $\alpha_S(M_Z)= 0.118$ \cite{PDG},  
after matching twice the QCD coupling constant at the quark-antiquark 
thresholds $\mu_{c \bar c}=2.4$ GeV and $\mu_{b\bar b}=10$ GeV. 

The distribution amplitude $\varphi_\pi$ can be expanded in terms of 
Gegenbauer polynomials $\Psi_n(u)=6u(1-u)C_{n}^{3/2}(2u-1)$.
Arguments based on conformal spin expansion \cite{BF} allow
us to neglect higher terms in this expansion. 
We adopt here the ansatz which consists of 
three terms, assuming that the terms $a_{n>4}$ are small 
\begin{eqnarray}  
\varphi_\pi(u,\mu_0)=\Psi_0(u) +a_2(\mu_0) \Psi_2(u) 
+a_4(\mu_0) \Psi_4(u).
\end{eqnarray}
The asymptotic distribution amplitude $\varphi_\pi(u)=6u(1-u)$ is unambiguously 
fixed \cite{BL}. 
The terms $n>0$ describe non-asymptotic corrections. 

We may consider 
two different approaches to confront the sum rule with 
the experimental data. 
The first possibility is to use the existing values 
for $a_2$, $a_4$ \cite{BF,CZ} in our sum rule 
and to compare the calculated form factor with the CLEO 
experimental data.
The second approach is to treat $a_2$, $a_4$ as unknown 
parameters and to extract them from the CLEO data. 
We shall discuss both possibilities, starting with the first one.

\subsection{The asymptotic, BF- and CZ- distribution amplitudes}

Previous extractions of  the parameters $a_2$, $a_4$ can 
be divided into three classes: 
\begin{itemize}
\item[1)]{Chernyak and Zhitnitsky (CZ) obtained 
the coefficients $a_2(\mu_0)=2/3$ and $a_4(\mu_0)=0$  \cite{CZ}
at the scale $\mu_0=0.5$ GeV.}
\item[2)]{Braun and Filyanov (BF)
extracted the coefficients 
$a_2(\mu_0)=0.44$ and $a_4(\mu_0)=0.25$  \cite{BF}
at the scale $\mu_0=1$ GeV.}
\item[3)]{Some groups argued that the wave 
function is very close to the asymptotic one, 
the coefficients are very close to $a_2=0$ and $a_4=0$ (see for example 
\cite{RadRus,RadMus,MikhRad,BakMikh,Belyaev}).}
\end{itemize}

The QCD evaluation of $a_2$ and $a_4$ at different scales $\mu$ gives
\begin{center}
\begin{tabular}{|c||c|c|c|c|c|c|c|c|}\hline
$\mu$& \multicolumn{2}{|c|}{$0.7\,\mbox{GeV}$}&
       \multicolumn{2}{|c|}{$1.0\,\mbox{GeV}$}&
       \multicolumn{2}{|c|}{$1.5\,\mbox{GeV}$}&
       \multicolumn{2}{|c|}{$2.4\,\mbox{GeV}$}\\\hline\hline
& $a_2$&$a_4$&$a_2$&$a_4$&$a_2$&$a_4$&$a_2$&$a_4$\\\hline
BF \cite{BF} & --- &--- & 0.44 & 0.25 & 0.33 & 0.17 & 0.28 & 0.13 \\\hline
CZ \cite{CZ} & 2/3 & 0    &--- & ---  & 0.33 & -0.008& 0.28& -0.009 \\\hline
\end{tabular}
\end{center}
Now we calculate the form factor using these values. 
We use the normalization point $\mu = 2.4\,\mbox{GeV}$, which   
corresponds to the virtuality of the photons in the central region 
of the experimental data. This scale is also used in many 
$B\to\pi$ calculations \cite{B2piQCD}. The lower scale 
$\mu=1.5$ GeV is taken in order to check the sensitivity of the results 
to the variation of $\mu$.

\begin{figure}[htb]
\center{\includegraphics[
        width=15cm,
        ]{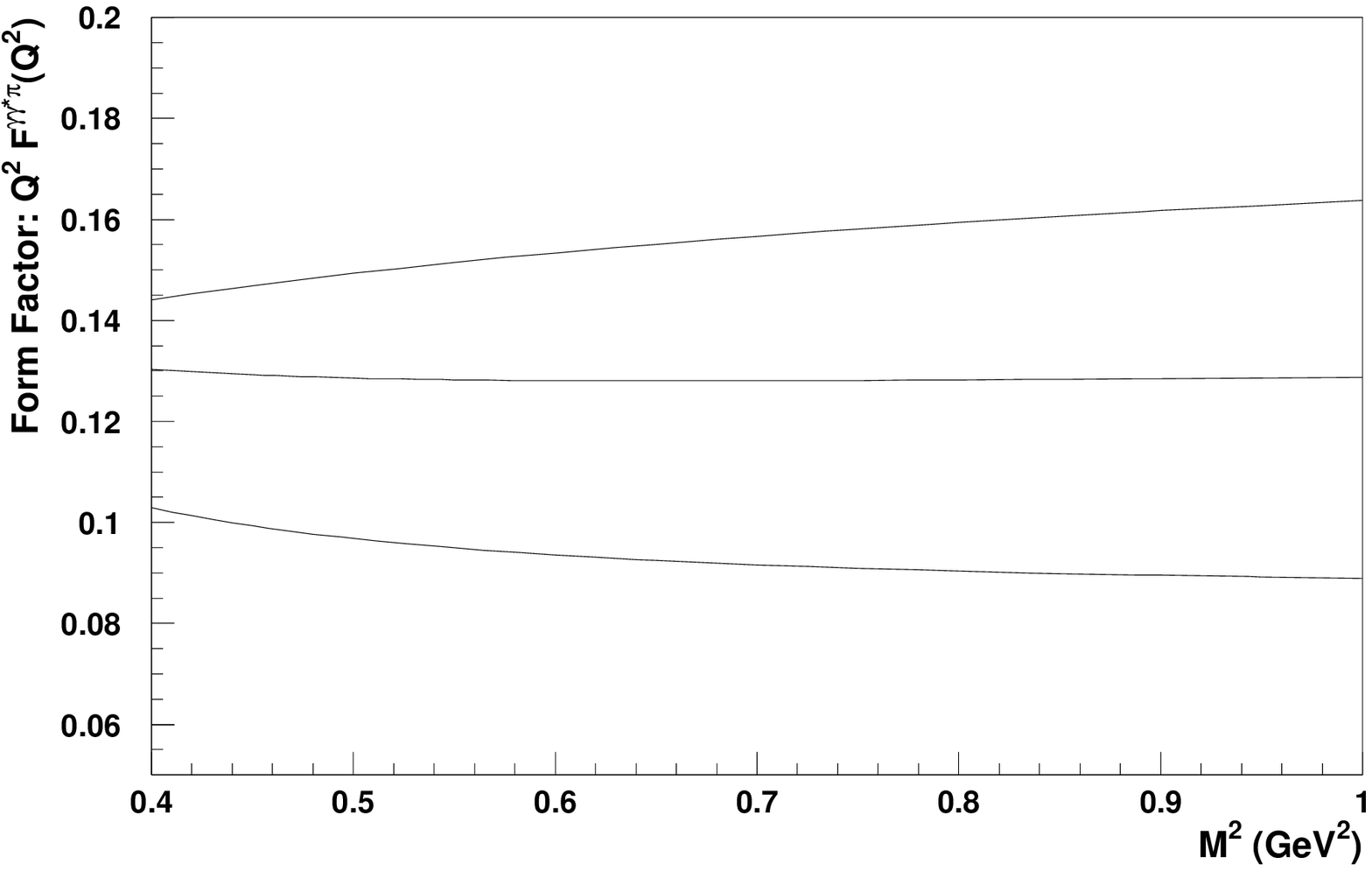}}
\parbox{12cm}{\caption{
The form factor $Q^2\,F^{\gamma\gamma^*\pi}(Q^2)$ as a function of the Borel
parameter $M^2$ for $Q^2=1,2,9\,\mbox{GeV}^2$ (from lower to upper line).}
\label{F:borel}}
\end{figure}

The sum rule depends on the Borel mass $M$ and 
the threshold energy $s_0$. 
The dependence on both parameters is small.
The variation of $s_0$ by $20 \%$ gives deviations 
in the form factor of less than $2\%$.
Fig.\ \ref{F:borel} 
shows the Borel mass dependence of the form factor at various
$Q^2\,\,(Q^2=-s_1)$. We note that the dependence is very small for
virtualities around
$Q^2\approx 2\div 3\,\mbox{GeV}^2$, 
where the experimental data is concentrated. 
For other $Q^2$ the Borel dependence is still moderate 
($\pm 2\div 4\%$), showing a good quality of the sum rule. 
 We use in the calculations $s_0=1.5\,\mbox{GeV}^2$ 
and $M^2 = 0.7\pm 0.2\,\mbox{GeV}$.

\begin{figure}[htb]
\center{\includegraphics[
        width=15cm,
        ]{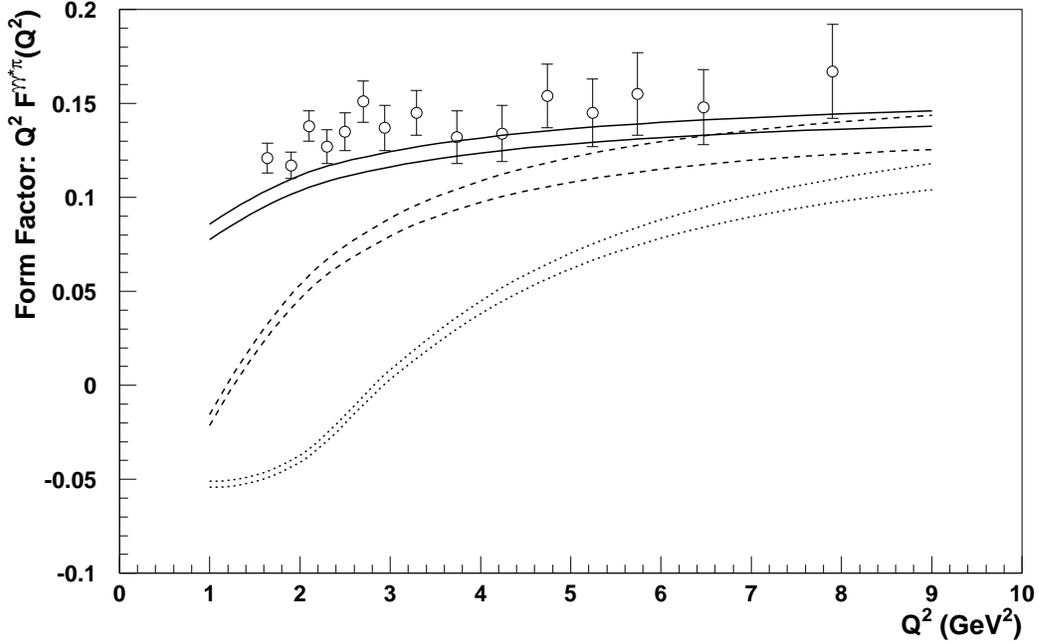}}
\parbox{12cm}{\caption{
The contributions of the asymptotic distribution amplitude
(solid line) and non-asymptotic terms of the first 
two Gegenbauer polynomials with $a_2=1$ in the whole $Q^2$-region 
(dashed line) and $a_4=1$ (dotted line) 
to the form factor $Q^2\,F^{\gamma\gamma^*\pi}(Q^2)$ 
as a function of $Q^2$ at the normalization point 
$\mu=2.4, 1.5$ GeV (upper, lower curves from a bunch of two curves 
correspondently)
The experimental data is taken from \cite{CLEO}.}
\label{F:plot1}}
\end{figure}

Now we are ready to compute the form factor using BF-, CZ- and 
asymptotic distribution amplitudes.
In Fig.\ \ref{F:plot1} we show the contributions of the asymptotic and
non-asymptotic parts of the distribution amplitude at $\mu=1.5$ and $2.4$ GeV; 
the parameters $a_2$ and $a_4$ are normalized 
to 1 in the whole region of $Q^2$. 
The twist-4 parameter $\delta^2(\mu)$ is fixed at $\mu=1$ GeV to be $0.2$
and scaled by the renormalization group equation with the 
one-loop anomalous dimension. 
We see that the asymptotic contribution 
is too small in order to fit experimental data.

\begin{figure}[htb]
\center{\includegraphics[
        width=15cm,
        ]{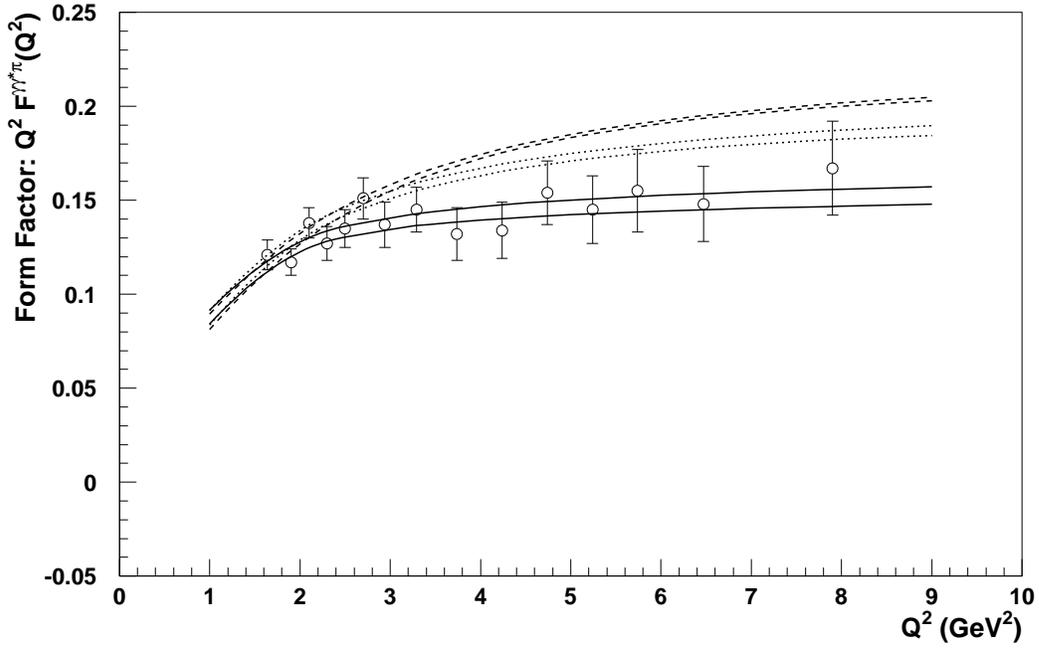}}
\parbox{12cm}{\caption{
The form factor  $Q^2\,F^{\gamma\gamma^*\pi}(Q^2)$ 
calculated with different distribution amplitudes:
the Braun-Filyanov (dashed lines), Chernyak-Zhitnitsky (dotted lines)
and our results extracted from CLEO data 
at $\mu=2.4$ GeV (upper lines from bunches of two lines) 
and  $\mu=1.5$ GeV (lower lines).
The experimental data is taken from \cite{CLEO}.}
\label{F:BFCZ}}
\end{figure}

The results with BF- and CZ- distribution amplitudes are presented in 
Fig.\ \ref{F:BFCZ}
and are too large to describe CLEO data.

\subsection{Extraction of the distribution amplitude from CLEO data}

We use a numerical nonlinear fit procedure to estimate best values for 
Gegenbauer coefficients.

 First, we study the ansatz for the pion distribution amplitude 
with only one non-asymptotic term $a_2$. 
Using this ansatz and comparing our results with CLEO results we obtain  
\begin{eqnarray}\label{result}
a_2 = 0.12 \pm 0.03\quad\mbox{at}\quad
\mu=2.4\quad\mbox{GeV}.
\end{eqnarray}
This result shows that the pion distribution amplitude is very close to
the asymptotic form. This conclusion is in agreement with recent analysis 
on electromagnetic pion form factor \cite{BKM} and also with 
results based on QCD sum rules presented in \cite{MikhRad,BakMikh,Belyaev}. 

Let us comment on the normalization scale dependence of our result, 
which enters through the QCD coupling constant 
and logarithms in the radiative correction. 
For the extraction of $a_2(\mu)$, 
we may choose any reasonable normalization point 
in the interval $1<\mu<3$ GeV. 
Different input values of $\mu$ will give different pairs 
of $a_2(\mu)$. We have checked that all extracted values of $a_2$ 
are in agreement with the renormalization group equation.
The details about the running of $a_n(\mu)$  with $\mu$ 
are discussed in Appendix B.

\begin{figure}[htb]
\center{\includegraphics[
        width=15cm,
        ]{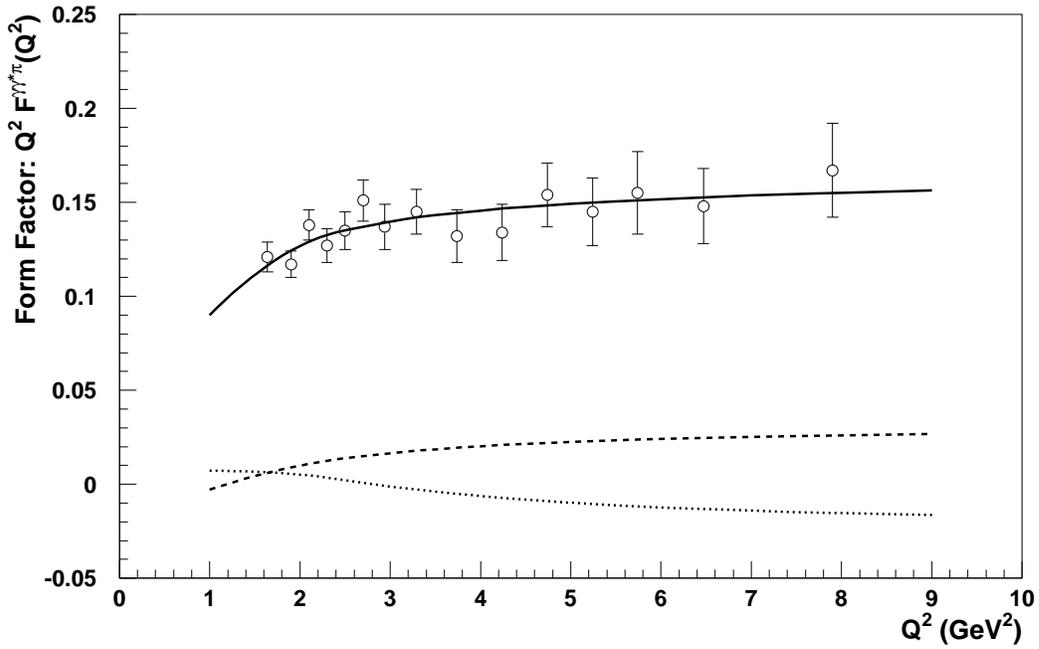}}
\parbox{12cm}{\caption{
The form factor  $Q^2\,F^{\gamma\gamma^*\pi}(Q^2)$ with the distribution amplitude 
extracted from CLEO data at $\mu=2.4\,\mbox{GeV}$ (solid lines).
The contribution of $a_2=0.19$ (dashed line) and $a_4=-0.14$ 
(dotted line) are the central values of our nonlinear fit.
The experimental data is  taken from \cite{CLEO}.}
\label{F:plot2}}

\end{figure}

We have also studied the ansatz for the pion distribution amplitude 
with two non-asymptotic terms $a_2$ and $a_4$.
 Using this ansatz and comparing our results with CLEO results we 
obtain:   
$a_2=0.19\quad \mbox{and}\quad a_4=-0.14\quad\mbox{at}
\quad\mu=2.4\,\mbox{GeV}$ as the fit parameters. 
The form factor calculated with the central values of the 
extracted parameters $a_2(2.4)$ and $a_4(2.4)$ is shown in 
Fig.\ \ref{F:plot2}.

\begin{figure}[htb]
\center{\includegraphics[
        width=15cm,
        ]{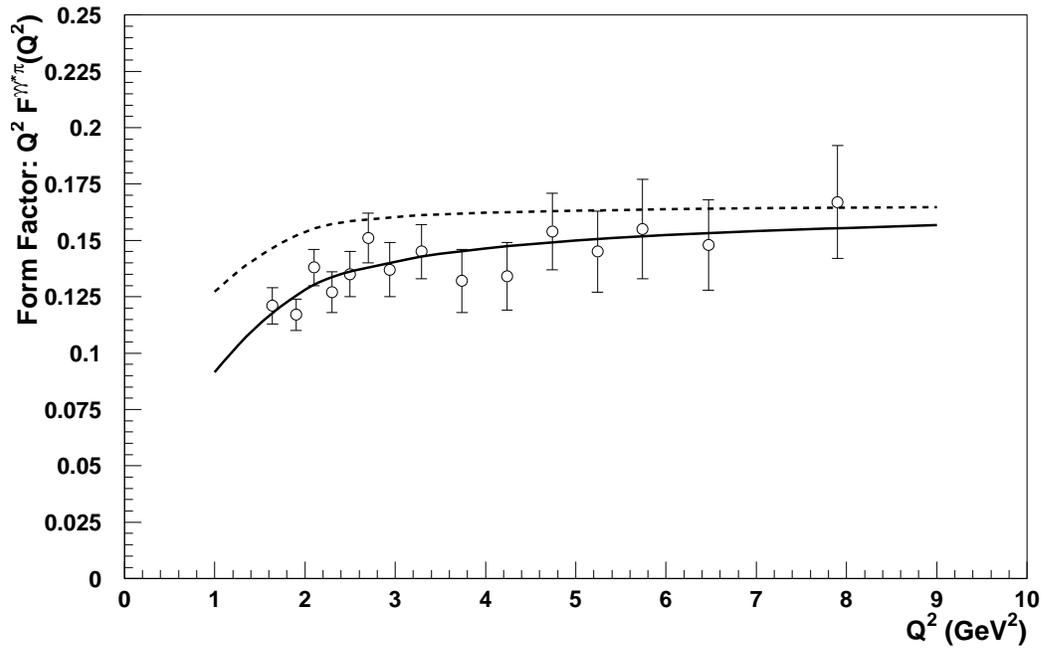}}
\parbox{12cm}{\caption{ The form factor  $Q^2\,F^{\gamma\gamma^*\pi}(Q^2)$ 
as a function of $Q^2$
at best extracted values of $a_2$ and $a_4$ at $\mu=2.4$ GeV
with (solid line) and without (dashed line) twist-4 contribution. 
The experimental data is taken from \cite{CLEO}.} 
\label{F:twist24}}
\end{figure}

Now let us discuss systematic uncertainties. 
There are many sources of systematic uncertainties.
One is the high order QCD perturbative corrections.  
Taking into account the size of one loop QCD correction 
($20\%$), we estimate two loop (and higher order correction) 
to be of order $0.2 \cdot 0.2 = 0.04 ( 4\% )$. 
Additionally, there are power corrections of twist higher than 4.
We assume them to be $20\%$ of the twist-4 contribution, 
which itself contributes about $20-30\%$ to the twist-2 term. 
The effect of the twist-4 contribution is shown in Fig.\ \ref{F:twist24}. 
We also adopt zero width approximation for $\rho$ and $\omega$.  
The uncertainty induced by this assumption is small, a few 
percent\footnote{In fact, at the $-s_1 > 4$ GeV$^2$ 
the resonance part in (8) (dual to the form factor $F^{\rho\pi}$) 
contributes less then $20-30\%$. The finite width 
of $\rho$ meson gives a deviation of order $10\%$ \cite{AK}, therefore, 
we are left with $2-3\% $ uncertainty in the region $-s_1> 4$ GeV$^2$.
It is worth to note that this energy region gives the same 
values for $a_2$, $a_4$ as the overall sample of data.},  
since the form factor  $F^{\rho\pi}$ does not appear in the final sum rule.
 The duality assumption introduces another source of uncertainties which is 
difficult to estimate. We assume that the variation of $M^2$ and $s_0$ gives
 us a rough idea about the size.

To fix statistical uncertainties we apply 
the fit procedure on a statistical set of data tables, where 
the experimental points are randomly displaced 
within the given errors \cite{CLEO}. 

\begin{figure}[htb]
\center{\includegraphics[
        width=15cm,
        ]{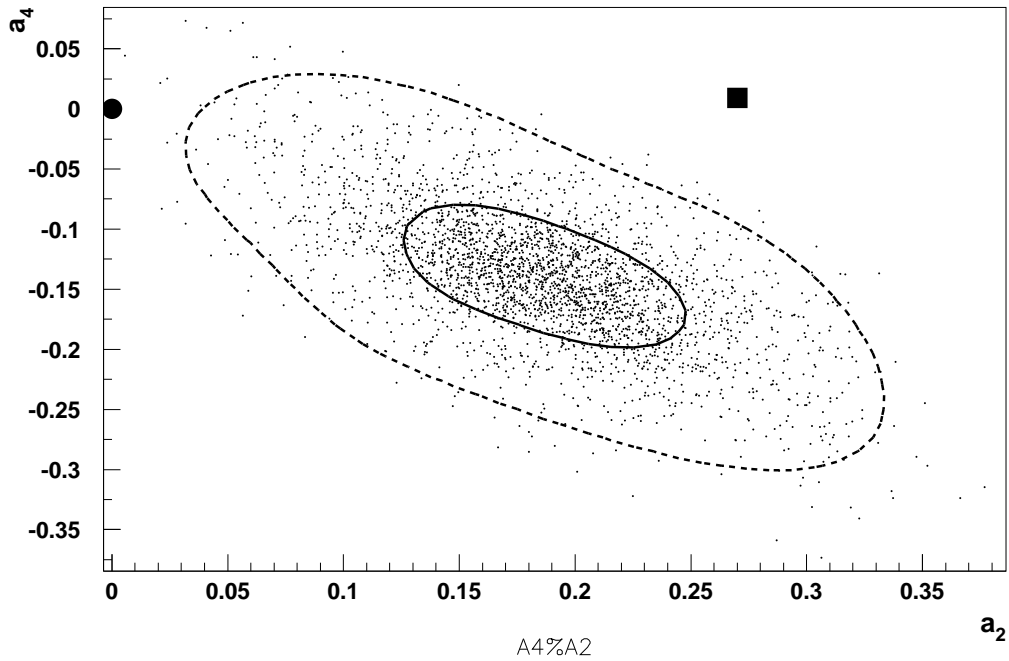}}
\parbox{12cm}{\caption{
The parameter space of ($a_2$, $a_4$) pairs extracted from 
3000 randomly chosen sets of data allowed by 
the experimental statistical uncertainties \cite{CLEO} as well as by 
the theoretical systematical uncertainties (\ref{Ffitted}).
Countor-lines show $68\%$ (solid line) and $95\%$ (dashed line)
confidential regions. Bold dots show the parameter pairs for asymptotic 
(circle) and Chernyak-Zhitnitsky (square) distribution amplitudes.}
\label{F:syserr}}
\end{figure}

Our results are presented in Fig.\ \ref{F:syserr}, 
where we show a parameter space of the twist-2 distribution amplitude,
which is obtained by comparing CLEO results with our sum rule.
 Besides of statistical uncertainties, we account for the 
possible theoretical uncertainties. 
We combine theoretical uncertainties together and write the theoretical form factor 
as (with 95\% CL):
\begin{eqnarray}\label{Ffitted}
F^{\pi\gamma\gamma^*}(s_1) = 
F_{As}^{tw-2}(s_1)(1\pm 5\%) + F_{As}^{tw-4}(s_1)(1\pm 20\%)
+ a_2^{fit} F_{a_2}(s_1) + a_4^{fit} F_{a_4}(s_1).     
\end{eqnarray}
The $F_{As}^{tw-2/4}$ denote the asymptotic contribution of twist-2/twist-4
distribution amplitudes, $F_{a_2}, F_{a_4}$ are the normalized 
contributions of the higher Gegenbauer polynomials. 

The width of the allowed region ( Fig.\ \ref{F:syserr})  
is due to experimental-statistical uncertainties, whereas 
the length is due to theoretical-systematical ones.

As we see form Fig.\ \ref{F:syserr}, theoretical-systematical
uncertainties are dominant over experimental statistical ones, 
giving a correlation between the parameters $a_2, a_4$.

\begin{figure}[htb]
\center{\includegraphics[
        width=15cm,
        ]{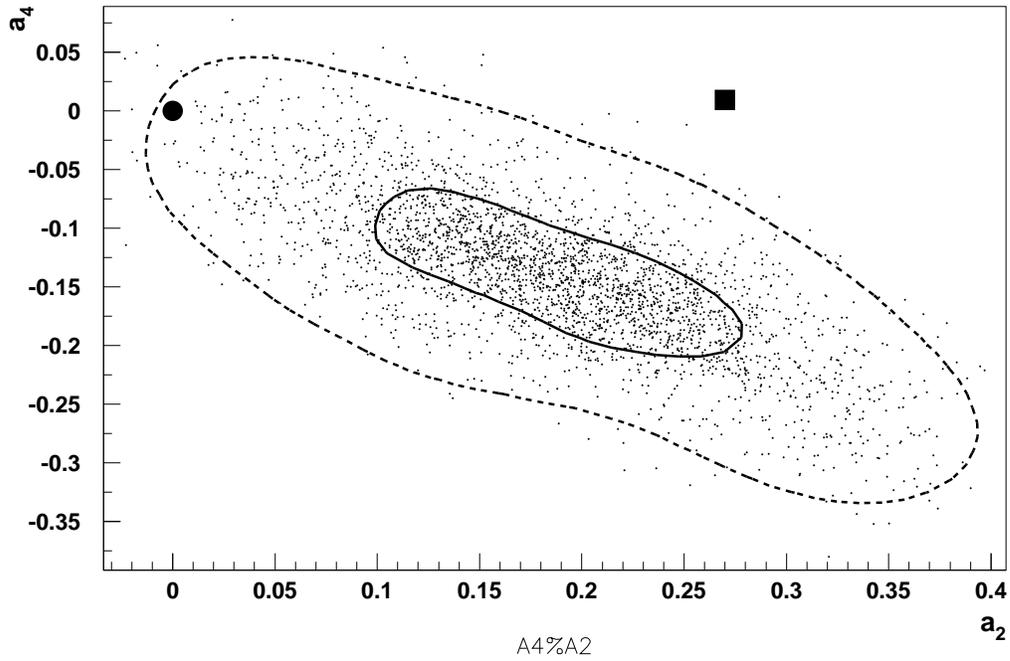}}
\parbox{12cm}{\caption{
The parameter space of ($a_2$, $a_4$) pairs extracted from 
3000 randomly chosen sets of data allowed by 
the experimental statistical 
uncertainties \cite{CLEO} as well as by the theoretical 
systematical uncertainties estimated {\em in a very conservative way}. 
Countor-lines show $68\%$ (solid line) and $95\%$ (dashed line)
confidential regions. Bold dots show the parameter pairs for asymptotic 
(circle) and Chernyak-Zhitnitsky (square) distribution amplitudes.}
\label{F:syserr2}}
\end{figure}

Fig.\ \ref{F:syserr2} shows a more conservative scenario where we assume an
uncertainty of $\pm 8\%$ in the twist-2 term and again $\pm 20\%$ for the 
twist-4 contribution. 

It is worth to mentioned that the allowed space of $a_2$ and $a_4$ does
not overlap with Braun-Filyanov and 
Chernyak-Zhitnitsky distribution amplitudes \cite{CZ,BF}. 
These distribution amplitudes are beyond the $95\%$ CL level 
region of the allowed parameter space for $a_2$, $a_4$. 
At the same time our region for $a_2,a_4$ is consistent with
results presented in \cite{MikhRad,BakMikh,Belyaev,BKM}, 
where it has been claimed that the pion distribution 
amplitude is very close to the asymptotic form.

The region presented in Fig.\ \ref{F:syserr2} does not favor any central
value. But in order to quantify the region, we may roughly identify 
following values for $a_2$ and $a_4$:
$a_2=0.19\pm 0.04\mbox{(stat.)}\pm 0.09\mbox{(syst.)}\,,
\quad a_4=-0.14\pm 0.03\mbox{(stat.)}\mp 0.09\mbox{(syst.)}$

We also observe that $F^{\pi \gamma\gamma^*}(Q^2)$ form factor is more
sensitive to the sum of $a_2$ and $a_4$, than to the  $a_2 - a_4$.
Really, we found that only one linear combination of $a_2$ and $a_4$ is
determined well, it is  $a_2+0.6a_4 = 0.11 \pm 0.03\quad\mbox{at}\quad
\mu=2.4\quad\mbox{GeV}$.

It would be very useful to study some observable, which is
sensitive to $a_2-a_4$, for example, the coupling constant $g_{\rho
\omega\pi}$. LCSR for this coupling is related to $\varphi(1/2)$, 
which is sensitive to the combination $a_2-a_4$. 

In general, the result for the model with one
non-asymptotic term (\ref{result}) is very reliable.   
At the same time, we have very useful constraint on the model with 
two non-asymptotic terms. Although, an additional constraint 
on $a_2-a_4$ would be extremely useful.

\section{Conclusions}

We have studied the pion form factor $F^{\pi \gamma\gamma^*}(Q^2)$ 
in the light-cone sum rule approach, including radiative 
corrections and higher twist effects.
 Comparing the results to the CLEO experimental data on 
$F^{\pi \gamma\gamma^*}(Q^2)$, we have extracted the pion 
distribution amplitude of twist-2. 
 The deviation of the distribution amplitude from the asymptotic 
one is small and is estimated to be $a_2(\mu) = 0.12 \pm 0.03 $ 
at $\mu=2.4$ GeV, in the model with one non-asymptotic term.   
This result on $a_2$ is in agreement with very recent 
analysis on electromagnetic pion form factor \cite{BKM}.
The ansatz with two non-asymptotic terms gives some region 
of $a_2$ and $a_4$, which is in qualitative agreement with results derived
in \cite{BakMikh,MikhRad,Belyaev,BKM}, 
but does not agree with CZ and BF models \cite{CZ,BF}.  

Since the theoretical systematical uncertainties 
are dominant over experimental statistical ones 
it will be very useful to estimate the power corrections 
and the higher order radiative corrections to the correlator  
in future. Also, it would be very useful to calculate LCSR for 
 some observable, which is sensitive to $a_2-a_4$, 
for example, the coupling constant $g_{\rho\omega\pi}$, and to
confront LCSR results with experiment.\\

{\bf Acknowledgments.}\\
We are grateful to A. Khodjamirian, R. R\"uckl, N. Harshman, F. von der
Pahlen for useful discussions and suggestions. 
We thank V. Savinov for providing us with 
details on the CLEO experimental data. This work is supported by 
the German Federal Ministry for Research and Technology 
(BMBF) under contract number 05 7WZ91P (0).

\newpage
\section{Appendix A}

To calculate the imaginary part, we use following identities:
\begin{eqnarray}\nonumber
\frac{1}{\pi}\mbox{Im}\,L_0 &=& \delta(u-u_0);\\\nonumber
\mbox{Re}\,L_0 &=& \mbox{P}\frac{1}{u-u_0};\\\nonumber
\vspace{3mm}
\frac{1}{\pi}\mbox{Im}\,L_1 &=&
-\Theta(u_0-u)\left(\frac{1}{u-u_0}\right)_+
   +\delta(u-u_0)\,\log|u_0|;\\\nonumber
\frac{1}{\pi}\mbox{Im}\,L_2 &=& -\Theta(u_0-u)\left(\frac{2\log|u_0-u|}
   {u-u_0}\right)_+ -
\delta(u-u_0)\left(\frac{\pi^2}{3}-\log^2|u_0|\right);
\\\nonumber
\frac{1}{\pi}\mbox{Im}\,l_1 &=& -\Theta(u_0);\\\nonumber
\mbox{Re}\,l_1 &=& \log|u_0|;\\\nonumber 
\frac{1}{\pi}\mbox{Im}\,l_2 &=& -2\Theta(u_0)\log(u_0);\\\nonumber
\mbox{Re}\,l_2 &=& \log^2|u_0| - \pi^2\Theta(u_0).\nonumber
\end{eqnarray}
\section{Appendix B}
The distribution amplitude $\varphi_\pi$ can be expanded in terms of 
Gegenbauer polynomials $\Psi_n(u)=6u(1-u)C_{n}^{3/2}(2u-1)$.
In NLO, the evolution of the distribution amplitude is given by \cite{KMR}:
\begin{eqnarray}
\label{phipi}
\varphi_\pi(u,\mu)= \sum\limits_{n} a_{n}(\mu_{0})
\exp\left(- \int\limits_{\alpha_{s}(\mu_{0})}^{\alpha_{s}(\mu)} d\alpha
 \frac{\gamma^{n}(\alpha)}{\beta(\alpha)} \right)
 \left( \Psi_{n}(u) + \frac{\alpha_{s}(\mu)}{4 \pi}
 \sum\limits_{k>n} d_{n}^{k}(\mu) \Psi_{k}(u) \right)
\end{eqnarray}
with $a_0=1$. The coefficients $d_n^k(\mu )$
are due to mixing effects, induced by the fact that 
the polynomials $\Psi_n(u)$ are
the eigenfunctions of the LO, but not of 
the NLO evolution kernel. The QCD 
beta-function $\beta$ \cite{PDG} 
and the anomalous dimension $\gamma^n$ of the 
$n$-th moment $a_n(\mu)$
of the distribution amplitude have to be taken in NLO.
Explicitly at $NLO$, the exponent in (\ref{phipi}) is  
\begin{eqnarray}
U(\mu,\mu_0)=\Big(\frac{\alpha_s(\mu)}{\alpha_s
(\mu_0)}\Big)^{\frac{\gamma^n_0}{2\beta_0}}
\Big(\frac{\beta_0+\beta_1\frac{\alpha_s(\mu)}{4\pi}}{\beta_0+
\beta_1\frac{\alpha_s(\mu_0)}{4\pi}} \Big)^{\frac{1}{2}
(\frac{\gamma^n_1}{\beta_1}-
\frac{\gamma^n_0}{\beta_0})}.
\end{eqnarray}
The anomalous dimensions \cite{anom} are 
\begin{eqnarray}
\gamma^n =\frac{\alpha_s}{4\pi}\gamma_{0}^n
+\left(\frac{\alpha_s}{4\pi}\right)^2\gamma_{1}^n
\label{gamman}
\end{eqnarray}
with 
\begin{eqnarray}
\gamma^{0}_0&=&0, \quad \quad
\quad \gamma^{0}_1=0~,
\nonumber
\\
\gamma^{2}_0&=&\frac{100}{9}, \quad \quad 
\gamma^{2}_1=\frac{34450}{243}-
\frac{830}{81}N_F,
\nonumber
\\
\gamma^{4}_0&=&\frac{728}{45}, \quad  \quad 
\gamma^{4}_1=\frac{662846}{3375}-
\frac{31132}{2025}N_F.
\end{eqnarray}
$N_F$ is a number of active flavours.
The beta-function coefficients are defined in a standard way \cite{PDG}.
The NLO mixing coefficients are \cite{KMR,Dittes}
\begin{eqnarray}
d_n^k(\mu) & = & \frac{M_{n k}}{\gamma_0^k-\gamma_0^n-2\beta_0}
\left( 1 - \left( \frac{\alpha_s(\mu)}{\alpha_s(\mu_0)} \right)^{\frac{\gamma_0
^k-\gamma_0^n-2\beta_0}{2\beta0}}
\right),
\end{eqnarray}
where the numerical values of the first few elements of the matrix 
$M_{n k}$ are 
\begin{eqnarray}
M_{0 2} = -11.2+1.73 N_F,~~ M_{0 4} = -1.41+0.565 N_F, ~~  
M_{2 4} = -22.0+1.65N_F.
\end{eqnarray}
The QCD evaluation of $a_2, a_4$ to the scale $\mu=2.4$ GeV gives 
$$a_2(2.4)=0.28,\qquad a_4(2.4)=0.13$$ for the Braun-Filyanov distribution amplitude 
and $$a_2(2.4)=0.28,\qquad a_4(2.4)=-0.009$$ 
for the  Chernyak-Zhitnitsky distribution amplitude. 

\newpage


\end{document}